\documentstyle[11pt,paspconf]{article}
\input psfig

\begin{document}

\title{\centering{The Structure of the Local Universe}\\
\centering{and}\\
\centering{the Coldness of the Cosmic Flow}\\}
\author{Rien van de Weygaert$^{a,1}$ and Yehuda Hoffman$^{b}$}
\affil{$^{a}$ Kapteyn Institute, University of Groningen, the Netherlands}
\vskip -0.2truecm
\affil{$^{b}$ Racah Institute of Physics, The Hebrew University, 
Jerusalem, Israel}


\altaffiltext{1}{Research Fellow of the Royal Netherlands Academy of 
Arts and Sciences}

\begin{abstract}
Unlike the substantial coherent bulk motion in which our local patch of 
the Cosmos is participating, the amplitude of the random motions around 
this large scale flow seems to be surprisingly low. Attempts to invoke  
global explanations to account for this coldness of the local cosmic 
velocity field have not yet been succesfull. Here we propose a different 
view on this cosmic dilemma, stressing the repercussions of our cosmic 
neighbourhood embodying a rather uncharacteristic region of the Cosmos. 
Suspended between two huge mass concentrations, the Great Attractor 
region and the Perseus-Pisces chain, we find ourselves in 
a region of relatively low density yet with a very strong tidal shear. 
By means of constrained realizations of our local Universe, based on 
Wiener-filtered reconstructions inferred from the Mark III catalogue of 
galaxy peculiar velocities, we show that indeed this configuration may 
induce locally cold regions. Hence, the coldness of the local flow 
may be a cosmic variance effect. 
\vskip -0.5truecm
\end{abstract}
\section{Introduction: Cosmic Migrations versus Local Chills} 
When speaking in terms of the motions of the objects populating the 
immediate vicinity of our Local Group, i.e. out to a distance of several 
tens of Megaparsec, our cosmic neighbourhood represents 
a rather chilly sector of the Universe. Rather than resembling a buzzing 
hive of galaxies rushing crisscross through space without any well-defined 
destination, we appear to be participating in a highly organized and coherent 
matter stream advancing towards a seemingly preordained direction. These  
streams are the instruments through which vast amounts of matter get 
channelled from their initial locations in the pristine and almost featureless 
primordial Universe towards the sites where matter accumulates in the 
process of building up the pronounced and complex patterns that nowadays 
we recognize in the large scale distribution of galaxies. 
\begin{figure}[t]
\vskip -1.5truecm
\centering\mbox{\psfig{figure=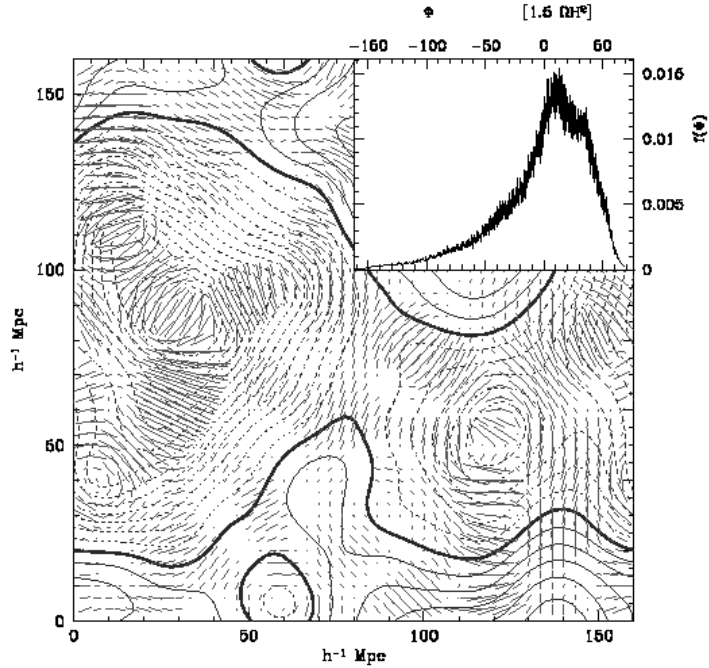,height=11.cm}}
\vskip -0.5truecm
\caption{The gravitational potential in the Local Universe: a ``bipolar 
world''. Contours: gravitational potential $\phi$. Bars: compressional 
component tidal shear. Insert: the pdf of $\phi$. Details: see text.}
\vskip -0.5truecm
\end{figure}
Within the gravitational instability scenario of structure formation, it are 
the same continuously waxing density excesses and depressions that 
induce these cosmic migration flows through their combined gravitational 
action. It is this intimate dynamical link between the distribution of matter, 
the induced cosmic flows, and the emergence of structure in the Universe 
that prompted substantial interest in the characteristics of 
the observed cosmic velocity field as useful fossil probes of the 
structure formation process. The amount of matter residing within 
the observed structures will be directly reflected in the corresponding 
matter streams. Hence, in a high-$\Omega$ Universe we expect the cosmic 
flows to involve higher velocities. This is true on any scale, whether it 
concerns the large scale bulk motions associated with the assembly of the 
characteristic foamlike patterns encountered on scales of tens of Megaparsec 
or structures on much smaller scales, whose dynamical timescales are 
so short that they have evolved to highly nonlinear stages in which the matter 
currents got (partially) ``thermalized'' and therefore lost the memory of 
their original state. 

Early assessment of the small-scale random motions of galaxies, estimated 
on the basis of pairwise velocity dispersions, revealed that locally the 
Universe is rather cold. While we participate in a bulk flow of 
$\approx 600 \hbox{km/s}$, the random velocities with respect 
to the mean flow are estimated to be in the range of a mere 
200--300 \hbox{km/s}. This low value of the velocity dispersion in 
combination with the pronounced 
structure displayed by the distribution of galaxies was in fact a strong 
argument for the latter being a biased tracer of the underlying matter 
distribution. The assumption of bias, in particular in the form of the 
(overly) simplistic linear bias factor $b$, would then imply the matter 
distribution not to have evolved as far as suggested by the pronounced 
nature of the galaxy distribution, and hence would be in agreement with 
the low value of the ``thermal'' motions in the local Universe.  
\section{Potential Eccentricities and Tidal Stresses}
In an attempt to interpret the significance of the coldness of the 
local cosmic flow, we postulate an alternative view. Rather than interpreting 
 the coldness of the flow as a property of the global Universe, we hold the 
view that the solution of the issue is contained in our rather atypical 
local cosmic vicinity. Within a distance of a few tens of Megaparsec 
we have not yet reached a fully representative volume of the Universe. This 
can be discerned rather straightforwardly from the local galaxy 
distribution. Even more explicit, however, is the situation when assessing 
the velocities of galaxies in our cosmic neighbourhood. These velocities 
reflect the underlying gravitational force and potential field, both 
having far larger coherence lengths than the density distribution. Hence, 
the volume probed by peculiar galaxy velocity surveys represents a 
rather restricted dynamical probe, unlikely to be anywhere near to fairly 
sampling that of the overall Universe. 

Meticulous analysis of the peculiar velocities of galaxies in the immediate 
cosmic neighbourhood have unveiled the nearby superstructures of the 
Great Attractor (GA) and the Perseus-Pisces supercluster (PP) as the 
dynamically dominating protagonists in the local cosmic tug of war. 
Although there are certainly several other contenders pulling their weight, 
be it in the form of nearby local structures or far-away monsters like the 
Shapley concentration, their contribution is unlikely to represent more than 
a moderate modification to the basic dynamical constellation set by the GA 
and the PP. Moreover, there does not appear to be any compelling evidence 
for the existence of dynamically influential mass concentrations beyond a 
distance of $\approx 150h^{-1}\hbox{Mpc}$.

A telling illustration of this preponderance of GA and PP is shown in 
the contour map of figure 1. It contains a reconstruction of the linear 
gravitational potential field (Gaussian scale $R_f = 5h^{-1}\,\hbox{Mpc}$) 
in our local Universe, within a region of size $160h^{-1}$ Mpc centered on our 
Local Group. Shown is a planar section approximately coinciding with the 
Supergalactic Plane. It is based on the set measured peculiar velocities 
of galaxies in the Mark III catalogue (Willick et al. 1997), processed by 
means of the Wiener filter reconstruction technique developed by 
Zaroubi et al. (1995). Evidently, the gravitational potential is dominated 
by two huge potential wells, the Great Attractor on the lefthand side and 
the Pisces-Perseus supercluster region on the righand side. Also the pdf of 
the gravitational potential (insert figure 1), displaying an atypical 
shoulder, corroborates the atypical nature of the local gravitational 
potential. Moreover, seemingly we find ourselves located right near the 
centre of a configuration strongly reminiscent of that of a canonical 
quadrupolar pattern, with two massive density enhancements along the 
horizontal axis and void regions concentrated around the perpendicular 
bisecting plane. The direct implication of the morphology depicted in 
figure 1 is that we are located near the saddle point of strong field of 
tidal shear. The red edges in figure 1, superposed on the potential contours 
and having a size and direction proportional to the strength of the 
compression along the indicated direction, represent the compressional 
component of this implied tidal force. Evidently, the tidal shear is very 
strong within the realm of the two huge matter concentrations where the 
density reaches high values. Very important to note, however, is that 
we also see that the tidal shear is indeed very strong near our own position, 
a region of rather modest density, due to the fact that we are located 
roughly halfway in between the Great Attractor and the Perseus-Pisces chain. 

Such a region of moderate to low local density in combination with a strong 
external tidal field may be expected to experience a different kinematical 
evolution from that of a similar isolated site. Its contraction will not 
only be dominated by its own selfgravity, external tidal forces of 
a similar order of magnitude will substantially shear the corresponding 
matter flows and lead to an anisotropic collapse. Although collapse may 
be accelerated along the compressional direction (Icke 1973), the 
shearing along the other directions may readily delay virialization 
and thus yield an ungenerically cold region.
\section{Cosmic Moulding}
To assess the viability of the heuristic picture sketched in the preceding 
section, we set out on an exploration of the kinematical evolution of 
a Local Group like region in an appropriate large-scale setting. The issue 
at stake involves the ``thermalization'' of a small-scale feature like our 
Local Group, definitively a nonlinear phenomenon, but also the influence of 
large-scale linearly or quasi-linearly evolving structures setting the 
external force field. No analytical approximations are known that would 
describe such situations to any satisfactory extent. This prompted our 
investigation by means of N-body simulations of the evolution of 
configurations resembling the local cosmic vicinity.
\begin{figure}[t]
\vspace{-1.5truecm}
\centering\mbox{\hspace{-0.5cm}\psfig{figure=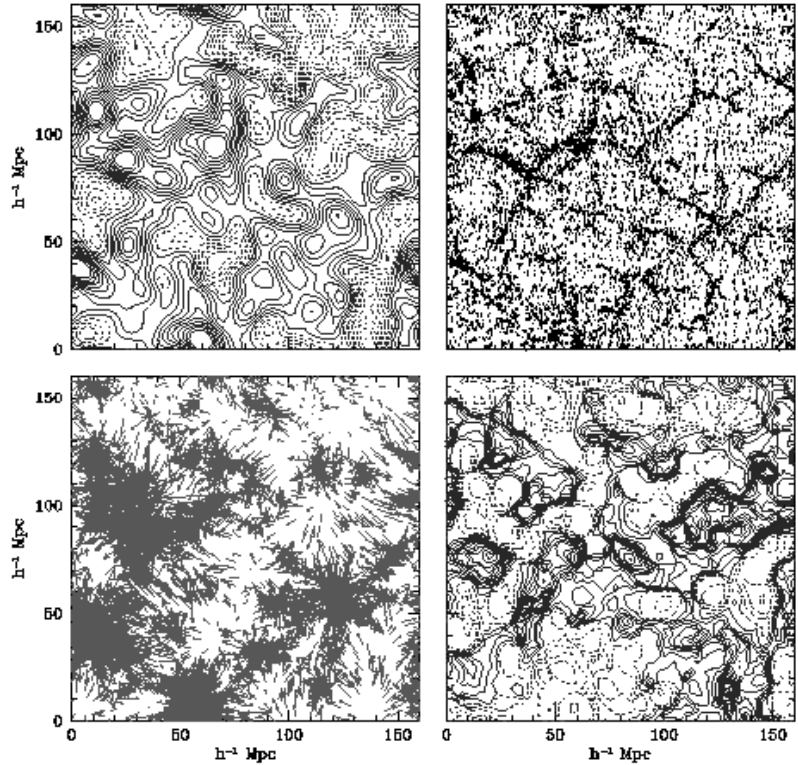,width=14.cm}}
\vskip -0.5truecm
\caption{Illustration of a computer realization/simulation of the local 
Universe. Top left: contour map initial density field. Top right: particle 
distribution at present epoch ($a_0$). Bottom left: particle peculiar 
velocities at $a_0$. Bottom right: contour map of local cosmic Mach number 
at $a_0$.}
\vskip -0.5truecm
\end{figure}
Of crucial importance to this study is the issue of an appropriate 
representation of 
the local Universe. We might restrict ourselves to some of the known 
properties of the Local Group, as for instance the density perturbation 
it represents, its peculiar velocity of $600\,\hbox{km/s}$, or maybe 
even some additional environmental requirements like the presence of 
a few nearby massive clusters. However, as indicated by e.g. Van de 
Weygaert \& Bertschinger (1996) such single localized constraints still 
leave ample freedom for the overall matter distribution. So much 
freedom that we cannot be guaranteed of actually assessing the 
appropriate cosmological situation. We would not be able to 
assure the correct spatial extent and structure of the large 
scale environment and consequently would also fail in having the 
correct temporal development of the local force fields. Because we argue 
that it is precisely the collusion between the specific localized conditions 
and the particular configuration of the the large-scale environment that may 
offer the explanation for the coldness of the local cosmic flow, 
we choose to set up optimally moulded initial conditions 
for our simulations instead of taking large random realizations and 
searching for objects that appear to be rather similar to our own 
Local Group. 

Focussing specifically on the dynamical and kinematical evolution of the local 
Universe, we invoke the observational information yielded by surveys 
of galaxy peculiar velocities, arguably the best available objective 
objective source on the mass distribution in the cosmic neighbourhood. 
On sufficiently large scales these velocities reflect directly 
the mass distribution, so that we can use them to distill an 
optimally significant reconstruction of the prevailing primordial 
linear density field in the local Cosmos. We achieve this by applying a 
Wiener filter algorithm to the sample of measured peculiar 
velocities of galaxies in the Mark III catalogue (Willick et al. 
1997), yielding a reconstruction that is optimal in terms of having a 
maximum signal/noise ratio (see Zaroubi etal. 1995). The Wiener filtered 
field in Figure 1 represents our cosmic environment on linear scales of 
$R_f > 5h^{-1}\,\hbox{Mpc}$. Such a reconstruction restricts itself to 
regions that are still within the linear regime, and whose statistical 
properties are still Gaussian. 

We discard further observational constraints on the small-scale 
clumpiness and motions in the local Universe. Instead, we generate 
and superpose several realizations of small-scale density and 
velocity fluctuations according to a specific power spectrum 
of fluctuations, with a global cosmological background specified by 
$H_0$ and $\Omega_0$, and with the small-scale noise being 
appropriately modulated by the large-scale Wiener filter 
reconstructed density field. To this end we invoke the technique 
of constrained random fields (see Hoffman \& Ribak 1991, van 
de Weygaert \& Bertschinger 1996), with the Wiener filtered 
field, ${\bf s}_{wn}=$ playing the role of ``mean field''
This setup allows us to test the likelihood of a cold local 
patch of the Universe embedded within a large scale environment 
reminiscent of the observed one in the local Universe. 
\section{Local Universe in a computer shell}
A particular constrained realization of a patch of the Universe 
resembling the primordial density field in our local cosmic neighbourhood 
is depicted in the upper lefthand frame of figure 2. It concerns 
a constrained realization of our Local Universe for the Standard 
Cold Dark matter scenario, with $\Omega_0=1.0$ and $H_0=50\,\hbox{km/s/Mpc}$. 
Its subsequent evolution is followed by means of the a P$^3$M N-body 
code.
The resulting distribution of the particles in a central slice through 
the simulations box is shown in the upper righthand frame of figure 2. 
Clearly recognizable are massive concentrations of matter at the 
locations where in the real Universe we observe the presence of the 
Great Attractor region (slightly ``north'' of the ``west'' direction) and 
the Perseus-Pisces region (slightly ``south'' of the ``east''). Interesting 
is to see how vast and extended these regions in fact are, certainly not to be 
identified with well-defined singular objects. 
\begin{figure}
\vspace{-1.5truecm}
\centering\mbox{\hspace{-1.0cm}\psfig{figure=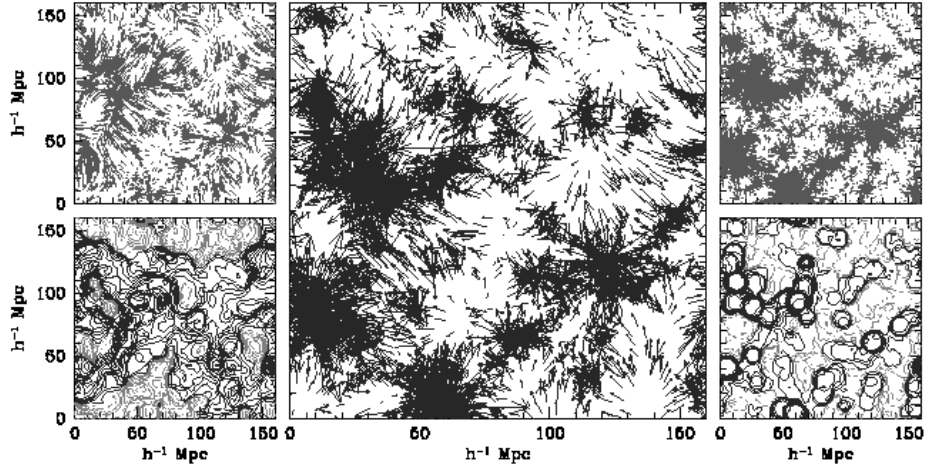,width=15.cm}}
\vskip -0.5truecm
\caption{Decomposition of the local velocity field (central frame) into 
the large-scale bulk velocity (lefthand frame) and residual small-scale 
``velocity dispersion'' (righthand frame).}
\vskip -0.5truecm
\end{figure}
\section{The local cosmic weather: Cool and Stormy}
The corresponding velocity field is shown in the lower lefthand frame. To 
appreciate the small-scale and large-scale contributions to the peculiar 
velocity field, in figure 3 we have decomposed the full velocity field 
(central frame) into the large-scale bulk flow ${\bf v}_{bulk}$ 
(lefthand, vectors in the top frame, amplitude contour plot in the lower 
frame) and the residual small-scale ``velocity dispersion'' 
${\bf v}_{\sigma}$ (for technical details see Van de Weygaert \& Hoffman 
1999). The decomposition criterion is set by top-hat filtering the 
velocity field at the scale of nonlinearity, $R_{TH}=8h^{-1}\hbox{Mpc}$. 
Note the striking matter displacement pattern revealed by the bulk flow 

field and its close relationship to the large-scale features emerging 
in the particle distribution (figure 2). Equally interesting is the fact 
that also the small-scale velocity dispersion field appears to bear the 
marks of underlying large-scale features: not only do we see substantial 
``thermal'' velocities at the sites of cluster 
concentratios, but we can also recognize sizeable small-scale velocities 
near the locations of filaments.
When we compare the contour plots of the bulk motion and the dispersion 
velocities, we can already discern the fact that while we -- located at the 
centre of the simulation box -- are embedded in a region with high bulk 
flow, evidently incited by the GA and the PP region, but that we also find 
ourselves in a region of exceptional low velocity dispersion. Indeed a 
telling reproduction of the observed ``coldness'' of the local cosmos. 
\begin{figure}[t]
\vskip -1.5truecm 
\centering\mbox{\hspace{-0.5cm}\psfig{figure=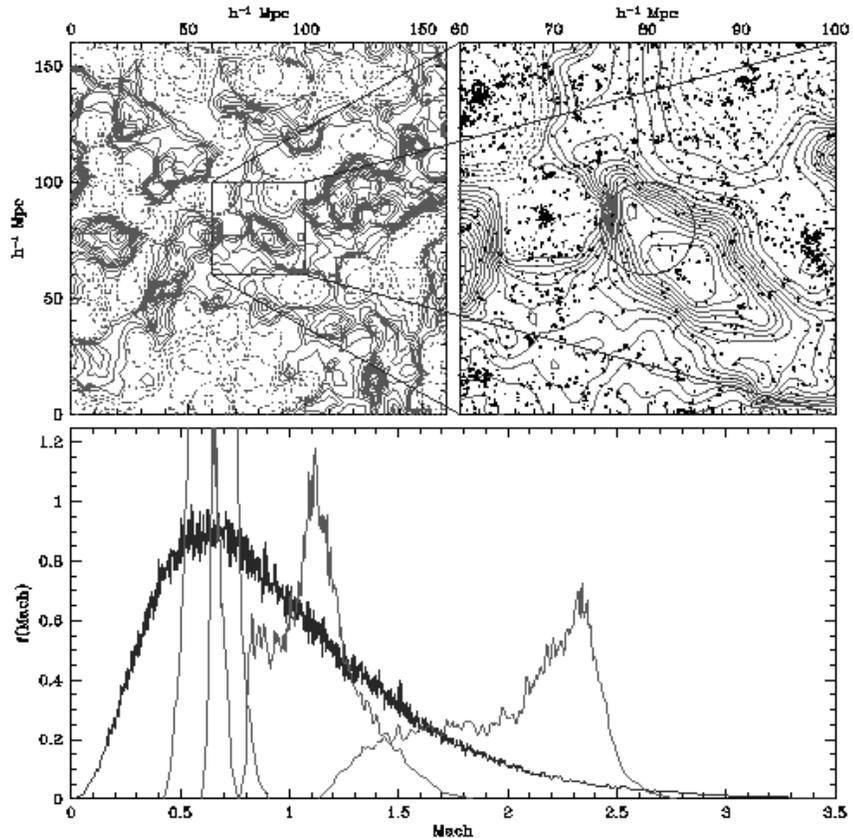,width=14.cm}}
\vskip -0.5truecm
\caption{Zooming in on the Mach number field in figure 2 (top frames). 
The global ${\cal M}$ pdf (solid), with a superposition of four 
locally determined pdf's (within sphere depicted in top righthand frame).}
\vskip -0.5truecm
\end{figure}
Assessing the relative large-scale and small-scale contributions 
to the velocity field, we quantify the ``coldness'' of the cosmic 
flow by means of the ratio between the local bulk flow velocities 
$|{\bf v}|_{bulk}$ and the ``dispersion'' velocity $\sigma({\bf v})$, 
\begin{equation}
{\cal M} \equiv {|{\bf v}|_{bulk} / \sigma({\bf v})}\,.
\label{eq:1}
\end{equation}
yielding a quantity whose cosmic average was introduced as 
``cosmic Mach number'' ${\cal M}$ by Ostriker \& Suto (1990). They propagated 
this quantity as a useful and complementary characterization of 
structure formation scenarios, quantifying the relative contributions 
of large and small scale matter perturbations with the great virtue 
of being insensitive to the amplitude of the power spectrum of 
density perturbations. We, however, are not so much interested in the 
cosmic average as well as in the spatial distribution and coherence of 
the point-to-point Mach number ${\cal M}({\bf x})$, and its relation 
to the underlying matter field. For the simulation mentioned above 
and illustrated in figure 2 and figure 3, we show the spatial 
structure of the Mach number field in the lower righthand frame of 
figure 2. A comparison with the corresponding density map reveals 
the interesting aspect of a large coherent band of high Mach number values 
running from the lower lefthand side to the upper righthand side of 
the simulation box, avoiding both the Great Attractor region and the 
Pisces-Perseus region, situated approximately in between those two 
complexes.
Superposed on this large-scale pattern are a plethora of small-scale 
features. For our purpose the most significant of these is 
the fact that we -- situated near the centre -- appear to be right near 
a towering peak of the Mach number distribution. Zooming in on the 
Mach number field we can readily appreciate this in figure 4, 
in the contour map frames. We should note that as the 
cosmic environment was specified on a scale of $R_g > 5h^{-1}\hbox{Mpc}$ we 
cannot really accurately pinpoint our own location to within a region 
of a similar size. In this respect it is very interesting to see a small 
Local Group size clump of particles near the peak of the Mach number 
distribution (top righthand frame fig. 4). It might indeed be the 
ultimate illustration of a Local Group like object, cold, relatively 
isolated, but member of a huge coherent complex. Even within a 
high-$\Omega$ Universe such a configuration may lead to an 
uncharacteristically cold situation. This can most clearly apprehended  
when comparing the global Mach number one-point probability distribution 
in the bottom frame of figure 4 with the distribution in limited 
central regions (sphere in the top frame). Superposed on the global 
pdf are distributions from central regions in four different 
realizations, the most shifted one corresponding to the simulation 
illustrated in the previous figures. Evidently, in two cases nothing 
exceptional is observed, but in two other cases we observe 
uncharacteristically ``cold'' environments. Hence, the atypical local 
velocity field realization may make us prone to infer flawed conclusions with 
respect to the global Universe, and we should take care to take into 
account the rather particular spatial local matter configuration in which 
we are embedded. 
\acknowledgments
We are grateful to A. Dekel for his permission to use the reconstructed 
density field based on the Mark III, and to B. Jones and E. Branchini 
for encouraging remarks.  

\end{document}